\documentclass[12pt,osajnl2,article,showpacs]{revtex4}  %% REVTeX 4.0

\begin{document}
\title{Haus/Gross-Pitaevskii equation for random lasers}

%% For REVTeX it is possible to automate superscript and e-mail callouts with the superscriptaddress option; see REVTeX4 documentation.

\author{Marco Leonetti,$^{1,*}$ and Claudio Conti,$^2$}
\address{$^1$ Dep. of Physics, University ``Sapienza'', Piazzale Aldo Moro 2, 00185 Roma, Italy}
\address{$^2$ CNR-ISC Institute for Complex Systems  Dep. of Physics, University ``Sapienza'', Piazzale Aldo Moro 2, 00185 Roma, Italy }
\address{$^*$Corresponding author: marcoleonetti1@gmail.com, current address: Photonic Crystal Group, ICMM,C.Sor Juana Inés de la Cruz, 3, Cantoblanco, 28049 Madrid, Spain.}

\begin{abstract}We report on experimental tests of the trend of random laser linewidth versus pumping power
as predicted by an Haus master equation that is formally identical to the one-dimensional Gross-Pitaevskii
equation in an harmonic potential. Experiments are done by employing picosecond pumped
dispersions of Titanium-dioxide particles in dye-doped methanol. The derivation of the master equations is also detailed and shown to be in agreement with experiments analytically predicting the value of the threshold linewidth.
\end{abstract}
%140.3410   Laser resonators;  140.3430   Laser theory; 140.3945   Microcavities; 140.3325   Laser coupling
% REPLACE WITH CORRECT OCIS CODES FOR YOUR ARTICLE
                          % NOTE: \ocis{} IS ALIASED TO \pacs{} BUT MUST
                          % FORMAT THE TERMS CORRECTLY FOR EACH JOURNAL

\maketitle %% null function with osajnl.sty

\section{Introduction}
Laser action is obtained by the simultaneous presence of gain due to stimulated emission and optical feedback. In a conventional laser these elements are embodied by an active medium placed between two mirrors that act as an optical resonator.
As predicted by Letokov \cite{Letokov} laserlike emission may be also obtained if  the resonator is replaced by a multiple scattering medium, (such as an ensemble of particles \cite{Lawandy} or atoms \cite{Kaiser}) which has the role of trapping light: if the volume of the inverted area  is sufficiently large to compensate the losses at the surface, a {\it random laser} (RL) is obtained.
This well known  phenomenon \cite{wiersma1} retains many features of standard lasers, such line narrowing at threshold, laser spiking \cite{Gouedard}, and coherence \cite{Zacharakis,Cao1}.

The main effect of multiple scattering is to increase the path length of photons inside the medium providing enhanced amplified spontaneous emission. If the scattering mean free path is much longer than the wavelength of light, the system may be described by using a diffusive model in which the ``photon particle'' is characterized by a linear increase of the mean square displacement with time, as in the standard of Brownian motion \cite{Einstein}. In this model, the energy flow inside the multiple scattering medium is treated by a continuity equation while neglecting phase and interference effects.
RL in the low scattering regime may be theoretically investigated by adding a gain term to the diffusive equation \cite{wiersma2}, which enables to predict the temporal shape of the emission or to study its  coherence properties \cite{florescu}.

At the end of last decade the presence of narrow intense spikes was discovered on random lasing spectra \cite{Cao1} in strongly scattering zinc oxide samples. The presence of these features can be explained as a signature of efficient resonant cavities, localized in a confined spatial region, in which the distribution of disorder determines the wavelength that feels higher amplification \cite{Laghendjik}. Although this point is still debated, and early works shows that the situation is more rich, random lasing may be seen as a superposition of electromagnetic modes put in oscillation in disordered fashion, with overlapping, finite spatial extent \cite{Angelani}. The nature of this kind of lasing modes is resonant and cannot be described by a diffusive model \cite{Tureci,Zaitsev09} that neglects interference effects.

In this manuscript we report on a model, originally introduced by the authors\cite{Conti}, in which RL action is attributed to many coupled modes with overlapping resonances, and this is taken as a starting point for deriving a nonlinear equation, which predicts the RL lineshape.
This theory is not limited by the diffusive approximation, which is not valid in the strongly scattering regime, and also not limited
to a specific dimensionality. Such an approach relies on a completely electromagnetic perspective, and allows (i) to derive closed-form analytical predictions, (ii) to rigourously define a threshold for the RL action and (iii) to predict the shape of the RL spectrum at various pumping intensities. The overall linewidth is described by an Haus Master equations \cite{Haus}
that is formally identical to a Gross-Pitaevskii equation \cite{Dalfovo,kivshar01}; its solution, either analytical (which is valid in proximity of the threshold) or numerical, provides a linewidth shape in quantitative agreement
with the experimental results.

In addition, the fact that the solution of a master equation, typically employed to describe ultra-short pulse generation, furnishes the description of laser emission in a disordered medium denotes that the latter can be interpreted as a coherent and collective emission of several electromagnetic resonances, eventually encompassing different degrees of localization, measured by the spreading in their life-time (temporal decay-constant) distribution.
This corresponds to the fact that all the resonances tend to vibrate with a deterministic phase-relation, i.e., through a spontaneous phase-locking mechanism. As also stated in the early thermodynamic treatments of lasers (see, e.g.,\cite{HakenBook}), such a process can be interpreted as a classical condensation process, that is a transition from a disordered ``thermal'' regime (all the modes oscillate independently) to a ``ferromagnetic-like'' regime (all the modes oscillates coherently). This links RL emission with recent investigation of condensation processes at a classical regime \cite{Picozzi}, with the remarkable difference that for RL the system is dissipative instead of Hamiltonian; in addition, this extends the thermodynamic approaches to lasers \cite{Weill05} to the case of disordered resonators\cite{leuzzi09}

This paper is organized as follows: in section II we discuss the current state of understanding concerning the
degree of localization of electromagnetic resonances in RL samples and we report on the
derivation of the Haus equation for RL and its theoretical predictions; in section III we report on the
comparison of the predicted linewidth with picosecond-pump RL, and conclusions are drawn in section IV.
\section{The Haus equation for Random Lasers}
We consider a disordered arrangement of dielectric scatterers in which the single particle, or interstices between them may act as optical cavities. Eigenmodes of a inhomogeneous dielectric material are eigensolutions of the Maxwell equation with a definite wavelength, spatial extent and lifetime (with reference to open systems).

A signature of presence of localized modes, has been firstly recognized in linear systems, by Wiersma et al. \cite{wiersma_sams}, as a decrement of the enhancement factor of backscattering cone in highly scattering media ($k\ell$ lower then 10,
with $l$ the transport mean free path  and $k$ the wavenumber) while Maret et al.\cite{Maret}
noticed the presence of light localization by measuring the time of flight in samples with  $k\ell$ 2.5.

When gain is introduced, light localizations become lasing modes that are spatially overlapping and compete for energy.
This results into spikes that are visible in the RL spectrum, as the  corresponding high Q cavities sustains a more efficient amplification due to lower losses with respect to extended modes.
Lagendijk et al, studied the spatial extent of lasing modes, \cite{Laghendjik} in gallium phosphide samples  ($k\ell\approx$ 6.4) embedded in an active medium pumped in a 3 $\mu m^2$ area. They retrieve a mode extension between 1 and 4 $\mu$m while numerical simulation \cite{Gentilini} results in a sub-micron localization length for comparable samples. Earlier studies demonstrate that coexistence between localized and extended modes in strongly scattering system plays a fundamental role  in the physics of RL \cite{Fallert} as they survive together in spatially extended random lasing with $k\ell \simeq4$ \cite{wiersma1}.
All these reported results, even if not conclusive, prove that even if Anderson localization is theoretically expected for  $k\ell\leq1$, a signature of the presence of localized modes can be found even for higher values. This means also that in condition of sufficiently strong scattering, diffusive approximation, that disregards any resonant behavior cannot be consistently applied.

In a previous paper\cite{Conti} the authors proposed an analytical model in which RL action is assumed to be sustained by a large number of electromagnetic resonances. Here we add more details on the theoretical part. Our picture is not affected the difference between localized modes and extended modes. Both types of modes have an eigenfrequency and can lase.

\begin{figure}[!ht]
\begin{center}
\includegraphics[width=8cm]{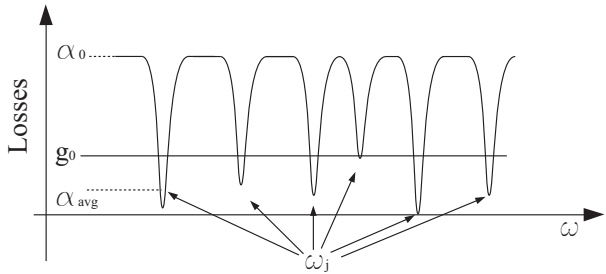}
\caption{Scheme of losses and gain profile as modeled by our theory.}
\label{fig: sceme}
\end{center}
\end{figure}
We start considering the spectral distribution the losses in highly scattering systems, these are expected to have a smooth profile interleaved by the high $Q$ resonant modes in the frequency domain. As schematically depicted in figure \ref{fig: sceme} the spectral profile of losses $\alpha (\omega)$ will appear like:
\begin{equation}\label{losses omega}
\alpha(w)=\alpha_0 -\sum_{j=1}^{N} \alpha_j(\omega-\omega_j)
\end{equation}
in which $\alpha_0$ is the average (nonresonant) value of losses and $\alpha_j(\omega-\omega_j)$ is a  sharply peaked (centered at $\omega_j)$ line shape corresponding to a localized mode j ($\alpha_j$ is centered at ($\omega=0$) for convenience). $\alpha_0$ is independent on the frequency $\omega$ due to the limited width of the spectral line.

The oscillation condition (gain must compensate losses) for the random lasing results to be:
\begin{equation} \label{oscillation condition}
g[\omega, A(\omega)]= \alpha(w) A(\omega)
\end{equation}
where A$(\omega)$ is the random lasing spectral content and $g[\omega, A(\omega)]$ is the spectral shape of gain, that nonlinearly depends on the whole spectral content because of the of the nonlinear susceptibility of the medium \cite{Haus,Lamb}:
\begin{eqnarray}\label{nonlinear susceptivity}
\nonumber
g[\omega, A(\omega)]= g_0\{(1-t_g^2 \omega ^2)A(\omega)  +\\
\int \int \int d\omega _1 d\omega _2 d\omega _3 \delta
(\omega +\omega_1 -\omega_2 -\omega_3) \chi (\omega_1;\omega_2 \omega_3)\\
A(\omega_1)^{\ast} A(\omega_2)A(\omega_3)\},
\nonumber
\end{eqnarray}
where $t_g$ is the lifetime of the gain bandwidth, and $g_0$ the linear gain coefficient.
By using (\ref{losses omega}) in the oscillation condition (\ref{oscillation condition}) we have
\begin{equation}\label{oscillation extended}
g[\omega, A(\omega)]=\alpha_0A(\omega) -\sum_{j=1}^{N} \alpha_j(\omega-\omega_j)A(\omega_j)
\end{equation}
in which we are allowed to substitute $A(\omega)$ with $A(\omega_j)$ in the second therm of the right side of the equation as, being $\alpha_j(\omega-\omega_j)$ much narrower of the spectrum $A(\omega)$, it will  ``probe'' only the resonance frequencies. As the number of the active localized modes in a macroscopic sample is enormous, we can suppose that the spectral distance of two contiguous resonances tends to zero, thus we  apply the continuous limit to equation (\ref{oscillation extended}):

\begin{equation}\label{oscillation cont lim}
g[\omega, A(\omega)]=\alpha_0A(\omega) -\int \alpha_{avg}(\omega-\Omega)A(\Omega)d\Omega.
\end{equation}

This equation can also be derived by assuming that all the modes are coupled with overlapping resonances, such that the Time Domain Coupled Mode \cite{Haus2} theory for the
generic mode $A_i$ is written as
\begin{equation}\label{tdcmt}
g[A_i]_i=\sum_j K_{ij}(\omega_i-\omega_j) A_j
\end{equation}
where $K$ is the coupling coefficients between two modes, that in general will depend on the distance between the resonance frequencies (the coupling will be vanishing as the spectral separation between modes increases). As the number of modes goes to infinity equation (\ref{oscillation cont lim}) is obtained, being
$\alpha_{avg}$ the average value of the couplings over all resonances and $\Omega$ takes the place of $\omega_j$.

By defining the Fourier transform as:
\begin{equation}\label{fourier_transf}
\mathcal{F}\left[a(\omega)\right]=\frac{1}{2 \pi}\int a(t)\exp(i\omega t)dt
\end{equation}
we may cast equation (\ref{oscillation cont lim}) in the time domain:
\begin{equation}\label{oscillation condition time domain}
g[t, a(t)]=[\alpha_0-\phi (t)]a(t)
\end{equation}
were by exploiting the convolution theorem, we substituted $\alpha_{avg}$, with his Fourier transform $\phi (t)$. $\alpha_{avg}$ is narrow with respect to the gain bandwidth; hence $\phi_L(t)$ can be expanded around $t=0$ with a parabolic function of time:
\begin{equation}\label{deeps_time_domain}
\phi_L(t)\cong(\alpha_0-\alpha_L)[1-(t/t_L)^2]
\end{equation}
where $\alpha_L$ is the average loss for the high-Q modes ($\alpha_0<\alpha_L$) and $t_L$ is their average lifetime.
Physically equation (\ref{deeps_time_domain}) has a simple interpretation: the various localized modes have a spread in their decay time distribution, this implies that at the beginning all the modes are put into oscillation and the average loss is high;
then short living (de-localized) modes or radiate out their energy or transmit it to long living modes, correspondingly
the average loss is reduced and the collective laser emission goes above threshold. Finally when
also the long living modes (that oscillate in phase during the emission) emit their radiation,
losses are increased again and the oscillation is below threshold.

The analytical form of gain in time domain is well known from the physics of mode locking \cite{Haus,JNKutz} and result from the Fourier transform of equation  (\ref{nonlinear susceptivity}):
\begin{equation}\label{gain time domain}
g[t, a(t)]=g_0\left[a(t) +t_g^2\frac{\mathrm{d}^2a(t)}{\mathrm{d}t^2}-\gamma_s|a(t)|^2 a(t)\right]
\end{equation}
where the  second therm inside the square parentheses results into the finite bandwidth of fluorescence, and the third  models the gain saturation. Lasing condition in the time domain turns out to be
\begin{eqnarray}\label{lasing condition time domain}
g_0\left[a(t) +t_g^2\frac{\mathrm{d}^2a(t)}{\mathrm{d}t^2}-\gamma_s|a(t)|^2a(t)\right]=\\\nonumber
\\
=[\alpha_0-(\alpha-\alpha_L)][1-(t/t_L)^2]a(t)\nonumber
\end{eqnarray}
By putting $a=a_0\varphi$ and $t=t_0\tau$, with
\begin{equation}
a_0^2=t_g\sqrt{\alpha_0-\alpha_L}
\end{equation}
\begin{equation}
t_0^2=t_g t_L\frac{\sqrt{g_0}}{\sqrt{\alpha_0-\alpha_L}}
\end{equation}
equation (\ref{lasing condition time domain}) can be cast, with the help of some algebra, in  a dimensionless form:
\begin{equation}\label{lasing condition dimensionless}
-\frac{\mathrm{d^2\varphi}}{\mathrm{d}\tau^2}+\tau^2\varphi+|\varphi|^2\varphi=E\varphi
\end{equation}
where the  ``nonlinear eigenvalue'' $E$ is given by
\begin{equation}\label{nonlinear eigenvalue}
E=\frac{t_L}{t_g}\frac{g_0-\alpha_L}{\sqrt{\alpha_0-\alpha_L}}=\frac{p-1}{\kappa\sqrt{p}}.
\end{equation}
Thus $E$ results to be determined by the adimensional pump energy as $p=g_0/\alpha_L$  and the constant $\kappa$ is defined as
\begin{equation}\label{kappa}
\kappa\equiv\frac{t_g}{t_L}\sqrt{(\frac{\alpha_0}{\alpha_L}-1)}.
\end{equation}
$\kappa$ is completely defined from the characteristics of the lasing material in fact $t_g$, $t_L$, $\alpha_L$, and $\alpha_0$ reflects  gain, resonance and scattering properties of the system.
Equation (\ref{lasing condition dimensionless}), takes in account for gain saturation, finite gain bandwidth, and the mode coupling due to overlapping resonances of the random lasing medium.
It has bell shaped solution for $E>1$ and this implies the presence of a sharp threshold for the laser action that may be defined as
\begin{equation}
p_{th}=1+\frac{\kappa^2}{2}+\kappa\frac{\sqrt{4+\kappa^2}}{2}.
\end{equation}
The RL spectral lineshape is found from the Fourier transform $\tilde \varphi(\tau)$ of the solution of Eq.(\ref{lasing condition dimensionless}) as:
\begin{equation}
S(\omega)=|A(\omega^2)|=\frac{t_g^2}{\gamma_s}|\tilde\varphi(\omega t_0)|^2
\label{spectrumRL}
\end{equation}.
\subsection{Generalized equation}
Equation (\ref{lasing condition dimensionless}) can be further generalized by accounting for higher order gain saturation,
indeed the correspoding time-domain gain is given by
\begin{equation}\label{gain time domain 2}
g[t, a(t)]=g_0\left[t_g^2\frac{\mathrm{d}^2a(t)}{\mathrm{d}t^2}+\frac{a(t)}{1+\gamma_s|a(t)|^2}\right],
\end{equation}
which reduces to (\ref{gain time domain}) in the small saturation limit.
Equation (\ref{lasing condition dimensionless}) becomes
\begin{equation}\label{lasing condition dimensionless 2}
-\frac{\mathrm{d^2\varphi}}{\mathrm{d}\tau^2}+\tau^2\varphi+\frac{1}{\epsilon}(1-\frac{1}{1+\epsilon |\varphi|^2})\varphi=E\varphi
\end{equation}
with $\epsilon=\gamma_s a_0^2$ a dimensionless parameter measuring gain saturation. As $\epsilon\rightarrow0$ equation (\ref{lasing condition dimensionless}) is obtained.

In the experiments reported below no significant discrepancy has been obtained when comparing the measured
quantities with equation (\ref{lasing condition dimensionless}) and (\ref{lasing condition dimensionless 2});
thus denoting the fact that the lowest order approximation for the gain saturation [Eq. (\ref{lasing condition dimensionless})] accurately describes
the experimentally accessible regime.
\subsection{Solution at threshold}
$\phi(\tau)$  and  its Fourier transform $S(\omega)$ (that is the intensity spectrum of the random laser) can be approximated by a gaussian near threshold,
indeed as $E\cong1$ it is (see, e.g.,\cite{kivshar01})
\begin{equation}
\varphi(\tau)\cong 2^{1/4} \sqrt{E-1} \exp(-\tau^2/2)
\end{equation}
which can be Fourier transformed and once recast in real-world units leads to
\begin{equation}
S(\omega)=\frac{t^2_g}{\sqrt{2}\pi\gamma_S}(E-1)\exp\left[\frac{-\omega^2}{8\pi^2W^{2}_{th}}\right]
\end{equation}
were the waist $W_{th}$ is
\begin{equation}
2\pi t_gW_{th}=\sqrt{\frac{\kappa}{2}}=\sqrt{\frac{t_g}{2t_L}\sqrt{\frac{\alpha_0}{\alpha_L}-1}}
\label{wthgauss}
\end{equation}
Note that a Gaussian lineshape was originally predicted by Lethovov \cite{Letokov}, in the framework of the diffusive approximation
for light propagation; in that case the width of the spectral waist was determined by Brownian motion of the
particles forming the scattering medium. Here our approach also holds well beyond the diffusive approximation, and no motion
is assumed for the disordered material in which the amplification is present. What is limiting the width of the Gaussian
spectrum is the distribution of decay times, and specifically the value coefficient of $\kappa$, which measures
(within numerical factors) the ratio between the spectral waist at threshold and the gain bandwidth ($\cong1/t_g$)
following equation (\ref{wthgauss}).

$t_L$ measures the average long-living modes decay time hence correlated value of losses is $\alpha_L\cong 1/t_L$. In addition $\alpha_0$ is the value of losses of the delocalized/diffusive modes, being
$l$ the transport mean free path and $v$ the energy transport velocity (which is of the order of $c/\bar n$
with $\bar n$ the average refractive index), it is $\alpha_0\cong v/l\cong D/l^2>>\alpha_L$ with $D$ the light diffusion constant
($D=v l/3$); furthermore for an almost localized regime $kl\cong1$,
hence $l\cong\lambda/2\pi$, which gives

\begin{equation}
\kappa^2\cong \frac{2\pi t_g}{\lambda} \sqrt{\frac{D}{t_L}},
\end{equation}
and
\begin{equation}\label{Diffusion_connection}
W_{th}\cong \sqrt{\frac{1}{4\pi t_g \lambda} \sqrt{\frac{D}{t_L}}}.
\label{wthd}
\end{equation}

Equation (\ref{wthd}) shows that the RL spectral waist in the localized regime decreases with the light diffusion constant
(which in finite-size real world system never vanishes at the localization), increases with the gain bandwidth, and is narrower
the longer is the spread of the decay time distribution or, equivalently, the longer the lifetimes of localized modes.

\subsection{Solution beyond threshold}
\begin{figure}[!ht]
\begin{center}
\includegraphics[width=8cm]{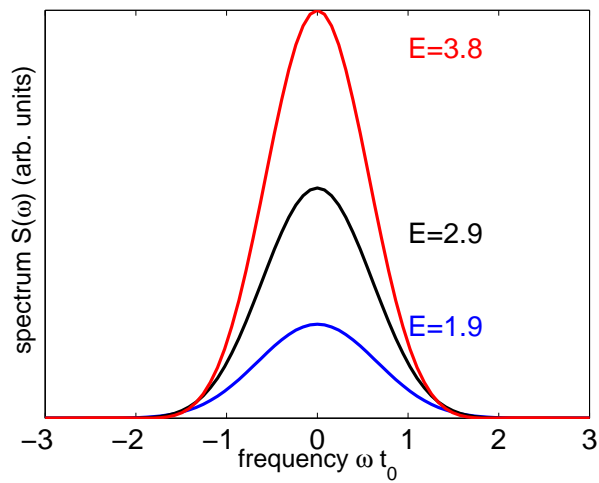}
\caption{(Color Online)Shape of the intensity spectrum for different values of the nonlinear eigenvalue E.}
\label{fig: theoretical spectras}
\end{center}
\end{figure}

\begin{figure}[!ht]
\begin{center}
\includegraphics[width=8cm]{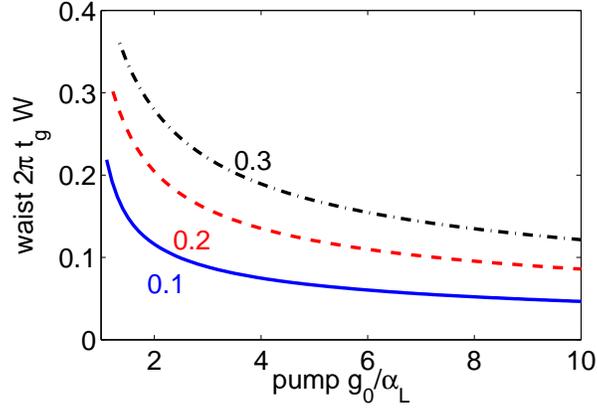}
\caption{ (Color Online) Waist (standard deviation) theoretically predicted for the RL spectrum in the high scattering regime.
The curves are shown for different values of the $\kappa$ parameter.}
\label{fig: theoretical FWHM}
\end{center}
\end{figure}

\begin{figure}[!ht]
\begin{center}
\includegraphics[width=8cm]{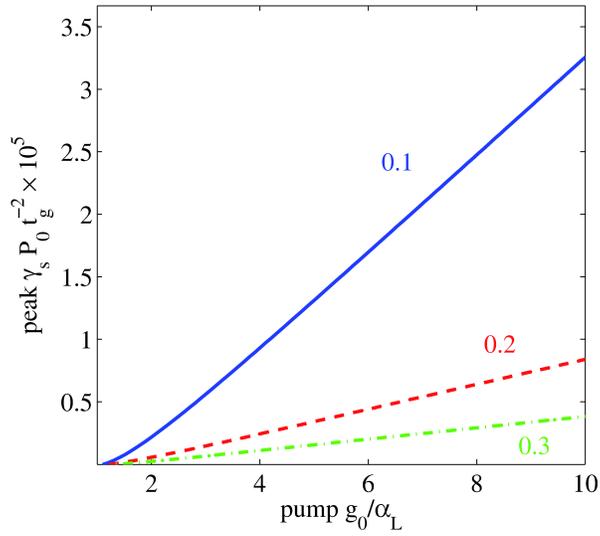}
\caption{(Color Online)
Peak intensity theoretically predicted for RL in the high scattering regime. The curves are shown for different values of the $\kappa$ parameter.}
\label{fig: theoretical peak}
\end{center}
\end{figure}

The predicted RL spectrum, as obtained after the
numerical solution of Eq.(\ref{lasing condition dimensionless}), is shown in figure \ref{fig: theoretical spectras} for an increasing nonlinear eigenvalue $E$ beyond the threshold.
Figure  \ref{fig: theoretical FWHM} and \ref{fig: theoretical peak} show respectively the waist and the peak of the spectrum
as functions of $E$, for different values of the $\kappa$ parameter.

Eq.(\ref{lasing condition dimensionless}) connects the random lasing spectra found by the nonlinear Schrodinger  equation (\ref{lasing condition dimensionless}) to the physics of Bose Einstein condensates \cite{Dalfovo}.
Indeed, Eq.(\ref{lasing condition dimensionless}) is formally identical
 to the bound state of a one-dimensional Gross-Pitaevskii equation that governs ultracold atoms
\cite{Dalfovo,kivshar01}.
The modulation of losses $\phi(\tau)$, which plays the role of the external potential $V(\vec{r})$, may be seen as a temporal trapping effect
that accounts for the existence of localized modes (low loss) that compete with extended ones.
In addition, this theoretical approach allows to obtain a spectral shape, and in particular the corresponding RL linewidth, as function
of the pumping energy density, thus furnishing an equivalent of the Schwalow-Townes \cite{Schawlow} law for RL.
\section{Experimental results}
\begin{figure}[!ht]
\begin{center}
\includegraphics[width=8cm]{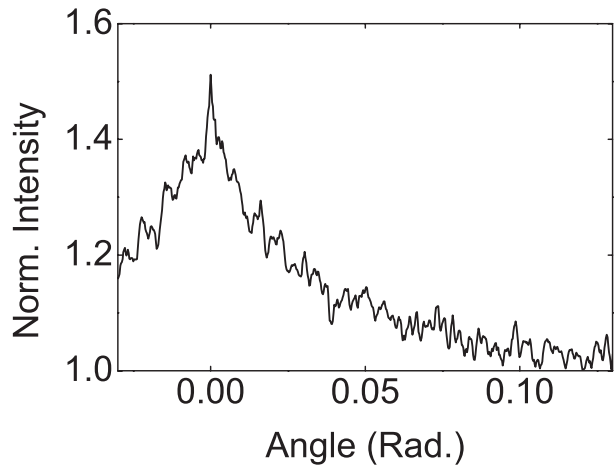}
\caption{Enhanced backscattering cone from disorderly arranged titanium dioxide particles ($\approx$ 300 nm diameter, 0.2 packing fraction) in methanol.}
\label{fig: backscattering}
\end{center}
\end{figure}

We used a colloidal dispersion of TiO$_2$  (Sachetleben Hombitan R611) particles in methanol doped by Rhodamine B (Sigma- Aldrich R6626, 10$^{-3}$ M).
We studied the sediment on the bottom of the couvette that deposes after half an hour from the preparation. The packing-fraction of the random lasing sample is 0.2 and its average refractive index $n_{av}$=1.5.
As the presence of the absorbing dyes makes impossible to perform elastic experiment, we measured mean free path by enhanced backscattering technique in a dye free solution containing titanium dioxide dispersed in methanol and NaCl (see figure \ref{fig: backscattering}). The presence of salt mimic the effect of the dyes on titanium dioxide (to screen Columbian interaction between particles) resulting in a sample  with a packing fraction closer to the active sample. We obtained  a value of $k\ell=8$ ($\ell=0.65$ $\mu$m ).
Pumping has been obtained by using a picosecond Nd:YAG frequency-doubled laser system (10Hz repetition rate, spot size 0.8 mm). Emission is retrieved by a fiber coupled spectrograph (Jobin Yvon, focal length 140 mm) and a thermoelectrically cooled CCD camera. The measured emission spectra rapidly shrinks above a threshold energy, and its waist qualitatively reproduces what predicted by equation (\ref{lasing condition dimensionless}).
To obtain a quantitative agreement between theory and experiments, we first measure
 1/t$_g\simeq$230nm  (in wavelength units)  by fitting the peak of the rhodamine fluorescence spectrum by a parabola
\begin{equation}\label{tgeq}
I(\omega)=I_0[1-(\omega-\omega_0)^2t_g^2]
\end{equation}
($\omega_0$ is the central wavelength of the fluorescence emission).

An estimate for the values of $\alpha_0$ and $\alpha_L$ are obtained by the properties of the system: $1/\alpha_0$ is the time needed to travel a mean free path:
\begin{equation}\label{alpha0}
\frac{1}{\alpha_0}=\frac{\ell}{c}n_{av}\approx3 fs
\end{equation}
while $1/\alpha_L$ may be found from the average of the inverse of the width of the random lasing spikes,
that are observed in the peak of the spectrum: $1/\alpha_L\approx2t_L\approx$10ps thus obtaining $\alpha_0/\alpha_L\approx$ 3300 and allowing to found an estimate of $\kappa\equiv\kappa_{th}\approx0.11$.
%
% The ratio between $t_g$ and $t_L$ may be found trough  the peaks linewidth ($\approx$ 0.5 nm) and the gain linewidth ($\approx$ 250 nm) resulting in\begin{equation}\label{TG_TL}
%\frac{t_g}{t_L}= 1/500
%\end{equation}

\begin{figure}[!ht]
\begin{center}
\includegraphics[width=8cm]{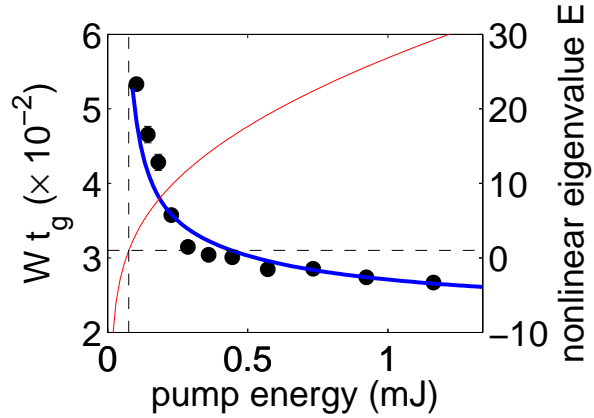}
\caption{(Color Online) Measured spectral linewidth Vs Energy (dots, left scale), the thick continuous line (left scale) is the best fit from the theory.
The right scale shows the trend of the adimensional nonlinear eigenvalue (thin line) versus the input energy,
as obtained from the fit.
\label{fig:fit_FW100ps}}
\end{center}
\end{figure}

To fit the data with our model, the nonlinear Schrodinger equation (\ref{lasing condition dimensionless}) is numerically solved to obtain the shape of the spectra for different values of the nonlinear eigenvalue $E$. The resulting relation between the waist $W$ and $E$ is approximated by a polynomial function $\mathcal{W}(E)$.
One has to find the relation between $E$ and the pumping energy of the laser $\overline{E}$.
The connection passes through the parameter $p$ defined in equation (\ref{nonlinear eigenvalue}). $p$ is proportional to the $g_0$ that is the linear gain, and correspondingly to the pumping energy $\overline{E}$.
We leave the parameter $\kappa$ and the constat of proportionality between $p$ and $\overline{E}$
\begin{equation}
p=\mathcal{C}\overline{E}
\end{equation}
as a free parameter of the fit.
Figure \ref{fig:fit_FW100ps} shows the normalized waist of the spectrum (calculated as the standard deviation) as function of the pumping intensity. From the fit we obtain an experimental value of $\kappa$ of 0.14 which is of the same order of magnitude of the estimated theoretical one above. The value of the threshold energy is found to be $\overline{E}_{th}=0.1$ mJ.
The analytically estimated spectral waist (after Eq.(\ref{wthgauss})) is hence given by $W_{th}=\sqrt{\kappa_{th}/2}/(2\pi t_g)\cong$10nm,
which is in quantitative agreement with the measured one.

\begin{figure}[!ht]
\begin{center}
\includegraphics[width=8cm]{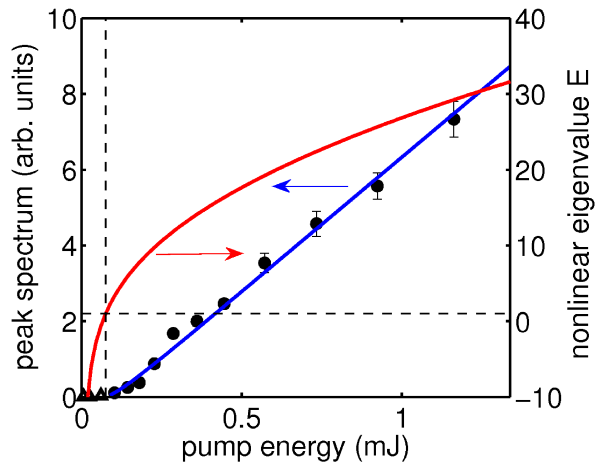}
\caption{(Color Online)
As in Fig.(\ref{fig:fit_FW100ps}) for the measured peak spectrum.
\label{fig:fit_peak100ps}}
\end{center}
\end{figure}

Similarly the predicted trend for the peak-spectrum also fit well with Eq.(\ref{spectrumRL}), as shown in Figure \ref{fig:fit_peak100ps}.
In this case the energy axis is the same as that determined for the waist in Fig.(\ref{fig:fit_FW100ps}) and a fitting scaling parameter is adopted for the vertical scale.

\section{Conclusions}
In conclusion we reported a detailed analysis concerning a novel theoretical model for random lasing in which light amplification is driven by a huge number of coupled resonant spatially localized modes.
Our approach does not need the diffusive approximation and results into a Gross-Pitaevskii equation, as derived by following the Haus theory of mode-locking, which plays the role of the Schwalow-Townes law for  RL and is in quantitative agreement with the experimental results, while also rigorously defining a threshold for the RL action.

With respect to \cite{Conti}, we clarify the role of losses in the time domain and the data fitting procedure. Moreover we derive equation (\ref{Diffusion_connection}) that connects diffusion constant to the lasing threshold.

Our results furnish novel insights on the nature of the random lasing phenomena, and open the way to further investigations on the phase-locking phenomena in disordered systems and generalized nonlinear equations for the corresponding emission spectral linewidth and temporal dynamics.

\section{Acknowledgments}
The research leading to these results has received funding from the European Research Council under the European Community Seventh Framework Program (FP7/2007-2013)/ERC grant agreement n.201766.

\end{document}